\documentclass[11pt,twoside]{article}
\usepackage{asp2006}
\usepackage{natbib,graphicx}

\begin{document}
\title{A Radio Study of 13 Powerful FRII Radio Galaxies}
\author{P. Kharb\altaffilmark{1}, C. P. O'Dea\altaffilmark{1}, 
S. A. Baum\altaffilmark{1}, R. A. Daly\altaffilmark{2}, 
M. P. Mory\altaffilmark{2},
M.~Donahue\altaffilmark{3}, E. J.~Guerra\altaffilmark{4}}
\altaffiltext{1}{\scriptsize Rochester Institute of 
Technology, Rochester, NY 14623, USA (Email: kharb@cis.rit.edu),
$^2$Penn State University, PA, USA, $^3$Michigan State University, MI, USA, 
$^4$Rowan University, NJ, USA}
\begin{abstract} 
We have observed a sample of 13 large, powerful Fanaroff-Riley type II radio 
galaxies with the Very Large Array (VLA) in multiple configurations 
and at multiple frequencies. We have combined our measurements of spectral
indices, rotation measures and structural parameters such as arm-length
ratios, axial ratios and misalignment angles, with similar data from the
literature and revisited some well-known radio galaxy correlations. 
\end{abstract}
\vspace{-0.8cm}
\section{The Radio Galaxy Sample}   
\vspace{-0.2cm}
In order to do a spectral aging analysis in the radio bridges of
Fanaroff-Riley-II (FRII) radio galaxies, we have observed a sample of 13 large, 
powerful FRII radio galaxies with the VLA using multiple configurations 
at 330~MHz, 1.4, 5 and 8~GHz \citep{Kharb06}. 
The sources span the redshift range of $0.4<z<1.6$ and 
have angular extents $>27\arcsec$. Using a 
large combined dataset comprising our radio galaxies and others from the
literature we have revisited some well known radio galaxy correlations, the
results of which are presented here. 

\vspace{-0.4cm}
\section{Results}
\vspace{-0.2cm}
Using the combined radio galaxy dataset we confirm that the
hotspot size $r_h$ is correlated with the total linear size $l$ of the
source and follows the relation $r_h\propto l^{0.7}$.
This result is consistent with a
self-similar model of a jet \citep[eg.,][]{Carvalho02b} propagating in 
a medium where the ambient
density $\rho_a$ falls off with distance from source $d$ as
$\rho_a\propto d^{-0.2}$. This could be due to the large jets spanning
hundreds of kiloparsecs in these sources, propagating mostly through
a roughly constant density intergalactic medium.
The hotspot spectral index is found to correlate with redshift and
follows the relation, $\alpha_{HS} \propto z^{0.4}$, which
is consistent with previous studies on radio galaxies 
\citep[eg.,][]{WellmanDaly97}.

The mean rotation measure (RM) of the radio lobes of the combined sample
is correlated with Galactic Latitude. The rotation measure dispersion,
on the other hand, is not. This suggests that the RM dispersion is 
probably caused by the source and/or its environment.
\begin{figure}
\centerline{
\includegraphics[width=6.9cm]{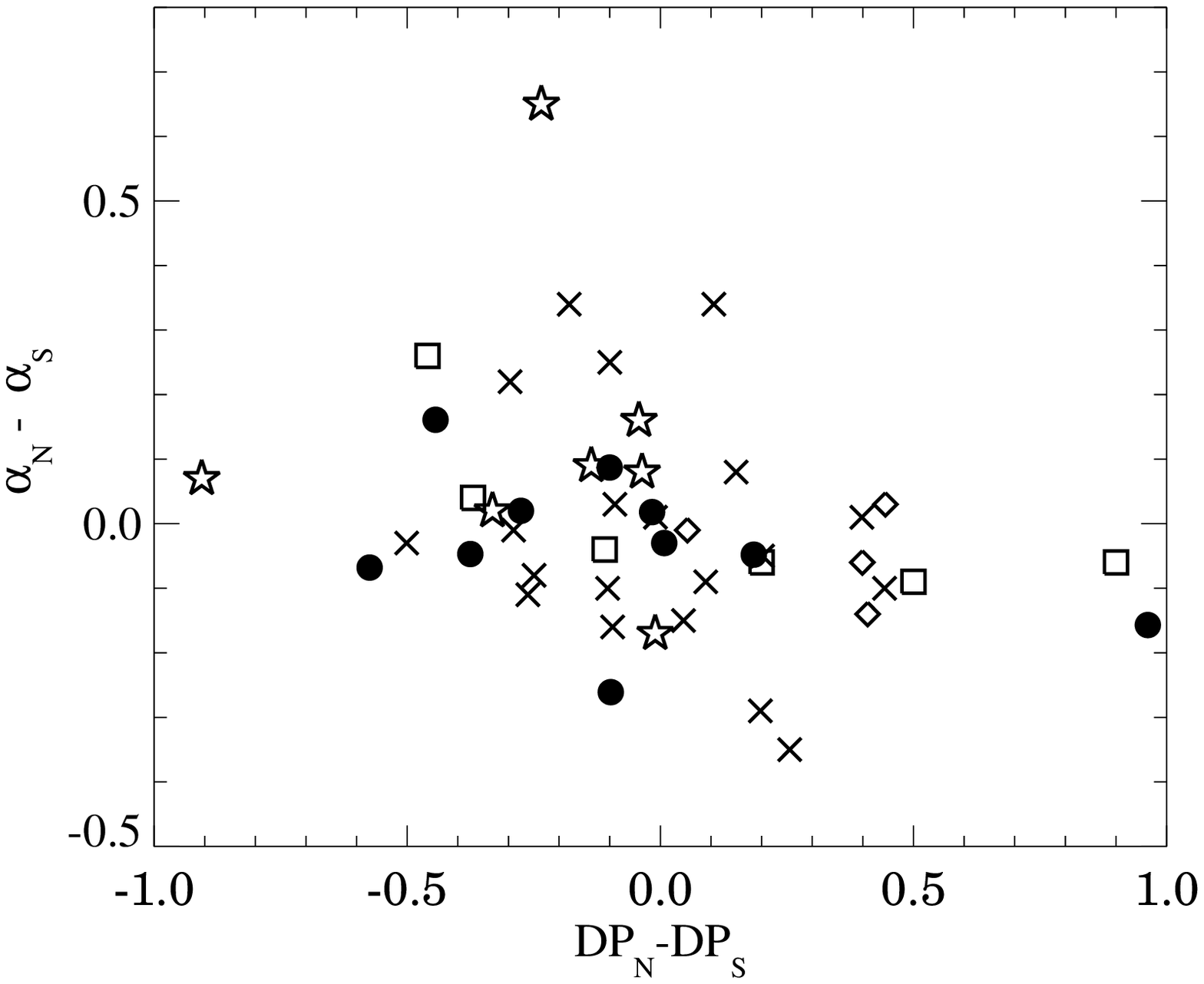}
\includegraphics[width=6.9cm]{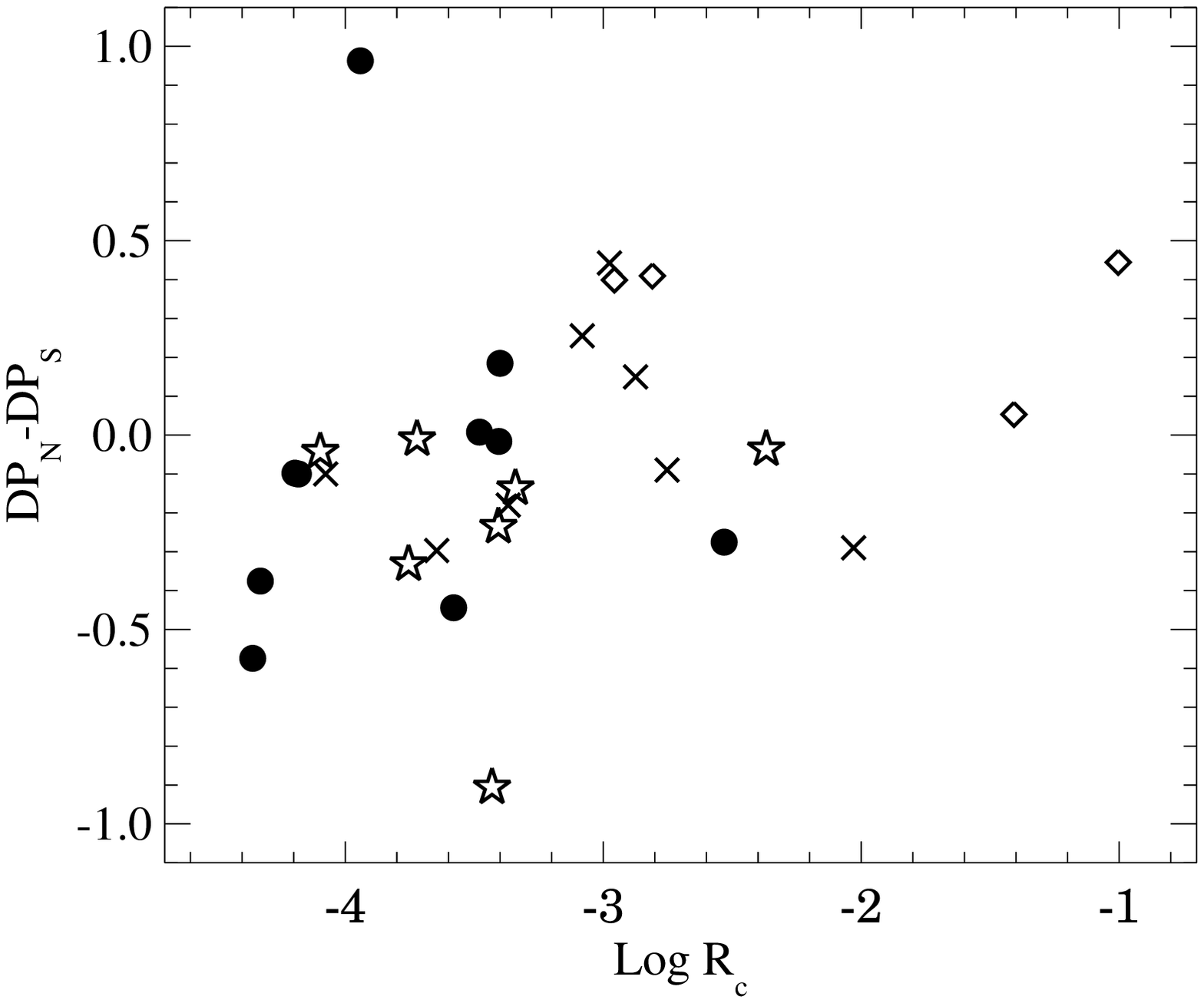}}
\vspace{-0.3cm}
\caption{The North-to-South-lobe depolarization (DP) difference versus
(Left) the difference in lobe spectral index (between 1.4 \& 5 GHz), 
and (Right) the K-corrected radio core prominence ($R_c$). 
DP is the ratio of the fractional polarization
in the lobe at 1.4 to 5 GHz; $R_c$ is the ratio of the radio core to lobe
flux density. Filled circles, stars, crosses, squares, and diamonds 
denote radio galaxies from
\citet{Kharb06,LiuPooleyA91,Goodlet04,Pedelty89}
and \citet{GarringtonConway91}, respectively. }
\end{figure}
We examined the Liu-Pooley correlation of lobe depolarization and spectral
index, with the combined dataset, and found
that the correlation is significant at the 99.99\% significance level
(Spearman rank test; Fig.1 Left). This is an improvement from
the original radio galaxy correlation (excluding quasars) which was observed
at the 80\% significance level. \citet{LiuPooleyA91} had concluded that
differences in the medium surrounding the two radio lobes influence both the
spectrum and the depolarization. A denser medium around one radio lobe would
result in greater confinement of the lobe, thereby decreasing the expansion
losses and increasing the radiative
losses, resulting in a steeper spectral index and greater depolarization.
\citet{McCarthy91} have indeed demonstrated that the emission-line gas is
intrinsically asymmetric in powerful radio sources. However, we find that
the depolarization does not seem to be correlated with the arm-length ratio or misalignment angle.

We find a weak correlation between lobe depolarization and radio core
prominence - which is a statistical indicator of beaming and therefore
orientation (Fig.1 Right). The weak correlation is 
consistent with the picture of these radio galaxies
lying largely in the plane of the sky. The lobe-to-lobe differences in 
spectral index however, do not correlate with the arm-length ratios, 
misalignment angles or radio core prominence. Further,
the arm-length ratios seem to be correlated with the misalignment angles but
anti-correlated with the axial ratios. This is suggestive of
environmental asymmetries close to the radio sources. Such asymmetries can
cause a variation in the outflow direction which can result in larger
misalignments between the two sides of the source. Variation in the jet direction can also result in
fatter radio lobes and lower axial ratios.

\acknowledgements 
This work was supported in part by U.S. National Science Foundation
under grant AST-0507465 (R.A.D.)


\end{document}